%% file: Neutrino_DM.tex
\documentclass[12pt]{article}
\usepackage{graphicx}

\newcommand\pubnumber{DPF2015-55}
\newcommand\pubdate{\today}

\def\montreal{Department of Physics Concordia University 7141 Sherbrooke W,\\ Montreal, Quebec,
Canada H4B 1R6}
\def\support{\footnote{Work supported by NSERC of Canada under SAP105354.}}

\def\Title#1{\begin{center} {\Large #1 } \end{center}}
\def\Author#1{\begin{center}{ \sc #1} \end{center}}
\def\Address#1{\begin{center}{ \it #1} \end{center}}

\newcommand\pubblock{\rightline{\begin{tabular}{l} \pubnumber\\
         \pubdate  \end{tabular}}}
\newenvironment{Abstract}{\begin{quotation}  }{\end{quotation}}
\newenvironment{Presented}{\begin{quotation} \begin{center} 
             Presented by SAHAR BAHRAMI at \end{center}\bigskip 
      \begin{center}\begin{large}}{\end{large}\end{center} \end{quotation}}

\input econfmacros.tex

\begin{document}
\begin{titlepage}
\pubblock

\vfill
\Title{Neutrino Dark Matter in the Higgs Triplet Model}
\vfill
\Author{ Sahar Bahrami and Mariana Frank\support}
\Address{\montreal}
\vfill
\begin{Abstract}
We analyze the effects of introducing vector-like leptons in the Higgs Triplet Model 
 providing the lightest vector-like neutrino as a Dark Matter candidate.  We explore the effect of the relic density constraint on the mass and Yukawa
coupling of dark matter, as well as calculate the cross sections for indirect and direct dark matter
detection. We show our model predictions for the neutrino and muon fluxes from the Sun, and the
restrictions they impose on the parameter space. We show that this model, with a restricted parameter space, 
is completely consistent with dark matter constraints, and indicate the resulting mass region for the dark matter.
\end{Abstract}
\vfill
\begin{Presented}
DPF 2015\\
The Meeting of the American Physical Society\\
Division of Particles and Fields\\
Ann Arbor, Michigan, August 4--8, 2015\\
\end{Presented}
\vfill
\end{titlepage}
\def\thefootnote{\fnsymbol{footnote}}
\setcounter{footnote}{0}

\section{Introduction}

We propose the Higgs Triplet Model (HTM) with vector-like leptons as a resolution to both neutrino masses and dark matter (DM) problems of the Standard Model (SM) \cite{Bahrami:2015mwa}. The resolution to neutrino masses is through the known see-saw mechanism, using an additional Higgs triplet representation.   The dark matter candidate is provided by a mostly singlet vector-like neutrino with small couplings to the $Z$ boson. A new parity symmetry, under which all new vector-like leptons are odd, prohibits mixing with the ordinary SM leptons.  Under this symmetry, the lightest odd particle 
  becomes stable on cosmological timescales, and is consistent with the DM candidate of the universe. We analyze the consequences of this scenario by requiring agreement with the relic density and non-collider experimental data, particularly with direct searches for  spin-independent (SI) or spin-dependent
(SD) interactions with target nuclei, with indirect dark matter searches, as well as with ultra-high energy neutrino experiments.  


\section{The HTM with Vector-like Leptons}
\label{sec:model}
The Lagrangian density for this model contains, in addition to the usual HTM $ \mathcal{L}_{\rm{kin}},~\mathcal{L}_{Y}$ and $V(\Phi,\Delta)$ terms for kinetic, Yukawa interaction for ordinary leptons, and potential terms, Yukawa interaction terms for the vector-like leptons $\mathcal{L}_{\rm VL}$ \cite{Bahrami:2015mwa}:
\begin{eqnarray}
\mathcal{L}_{\rm{HTM}}= \mathcal{L}_{\rm{kin}}+\mathcal{L}_{Y}+\mathcal{L}_{\rm VL}-V(\Phi,\Delta). 
\end{eqnarray}
The vector-like leptons in the model, with representations and quantum number are given in the Table below, and their interactions are given by $\mathcal{L}_{\rm VL}$: \begin{table}[htbp]
  \begin{center}
 \small
 \begin{tabular*}{0.99\textwidth}{@{\extracolsep{\fill}} c| ccccccc}
 \hline\hline
	Name &${\cal L^\prime}_L$ &${\cal L^{\prime \prime}}_R$ &${e^\prime}_R$ &${e^{\prime \prime}}_L$ &${\cal \nu^\prime}_R$ &${\cal \nu^{\prime \prime}}_L$ 
	\\
  \hline
  Quantum Number &$(\mathbf{1}, \mathbf{2}, -1/2)$ &$(\mathbf{1}, \mathbf{2}, -1/2)$ &$(\mathbf{1}, \mathbf{1}, -1)$ &$(\mathbf{1}, \mathbf{1}, -1)$ &$(\mathbf{1}, \mathbf{1}, 0)$ &$(\mathbf{1}, \mathbf{1}, 0)$  \\
      \hline
    \hline
   \end{tabular*}
\end{center}
 \end{table}
 \vskip-0.5in
\begin{eqnarray}
\mathcal{L}_{\rm VL}&=&-\big [M_{L}{\bar L}_L^\prime L_R^{\prime \prime}+M_{E}{\bar e}_R^{\prime} e_L^{\prime \prime}+ M_{\nu}{\bar \nu}_R^{ \prime} \nu_L^{\prime \prime}+\frac12 M_\nu^\prime {\overline \nu^{\prime c}_R} \nu_R^\prime + \frac12 M_\nu^{\prime \prime} {\overline \nu^{\prime \prime c}_L} \nu_L^{\prime \prime}+h_E^\prime ({\bar L}_L^\prime \Phi )e_R^\prime
 \nonumber \\
&& +h_E^{\prime \prime} ({\bar L}_R^{\prime \prime} \Phi )e_L^{\prime \prime}+h_\nu^\prime ({\bar L}_L^\prime \tau \Phi^\dagger )\nu_R^\prime +h_\nu^{\prime \prime} ({\bar L}_R^{\prime \prime} \tau \Phi^\dagger )\nu_L^{\prime \prime} + h^\prime_{ij}\overline{L_L^{\prime\, c}}i\tau_2\Delta L_L^{\prime\, }
+h^{\prime \prime}_{ij}\overline{L_R^{\prime \prime \, c}}i\tau_2\Delta L_R^{\prime\prime } \nonumber \\
&&+  \lambda_E^i ({\bar L}_L^\prime \Phi )e_R^i +  \lambda_L^i ({\bar L}_L^i \Phi )e_R^\prime + \lambda^\prime_{ij}\overline{L_L^{ ic}}i\tau_2\Delta L_L^{\prime}
+ \lambda^{\prime \prime}_{ij}\overline{L_R^{ ic}}i\tau_2\Delta L_R^{\prime \prime}   +\rm{h.c.} \big].~~~~ \label{vl_lgr}
\end{eqnarray}
Taking $h_\nu^\prime \ne 0$, $M_\nu=0$, but $h_{\nu}^{\prime \prime}=0$, the lightest neutral eigenvalue provides a single DM candidate of mass $M_{nu_1}$ and a single Yukawa coupling which are parameters we vary to obtain consistency with the experiment.  
\section{Dark Matter Relic Density}
\label{sec:RD}
%
The relic density  $\Omega_{DM}$ of non-baryonic DM in the energy-matter of the universe from cosmological data must be consistent with any model analyses.  We calculate it, restricted to $2\sigma$ allowed range  $0.1144 \leq \Omega_{DM} h^2 \leq 0.1252$, as constrained by WMAP \cite{Komatsu:2010fb} and PLANCK \cite{Ade:2013zuv} and present it below in Fig. \ref{RD}, 
 as a function of the DM mass and  Yukawa coupling $h_\nu^\prime$.  The two dips at $M_{DM} \sim 45$  GeV and $ \sim 62$ GeV correspond to resonant annihilation into $Z$ bosons and Higgs boson $h$.  Relic density constraints restrict the dark matter mass to be heavier than 23 GeV and lighter than 103 GeV and independent of any other parameters, to which it is insensitive.  

\begin{figure}[htbp]
\begin{center}
\hspace*{-0.5cm}
    \includegraphics[width=2.3in,height=2in]{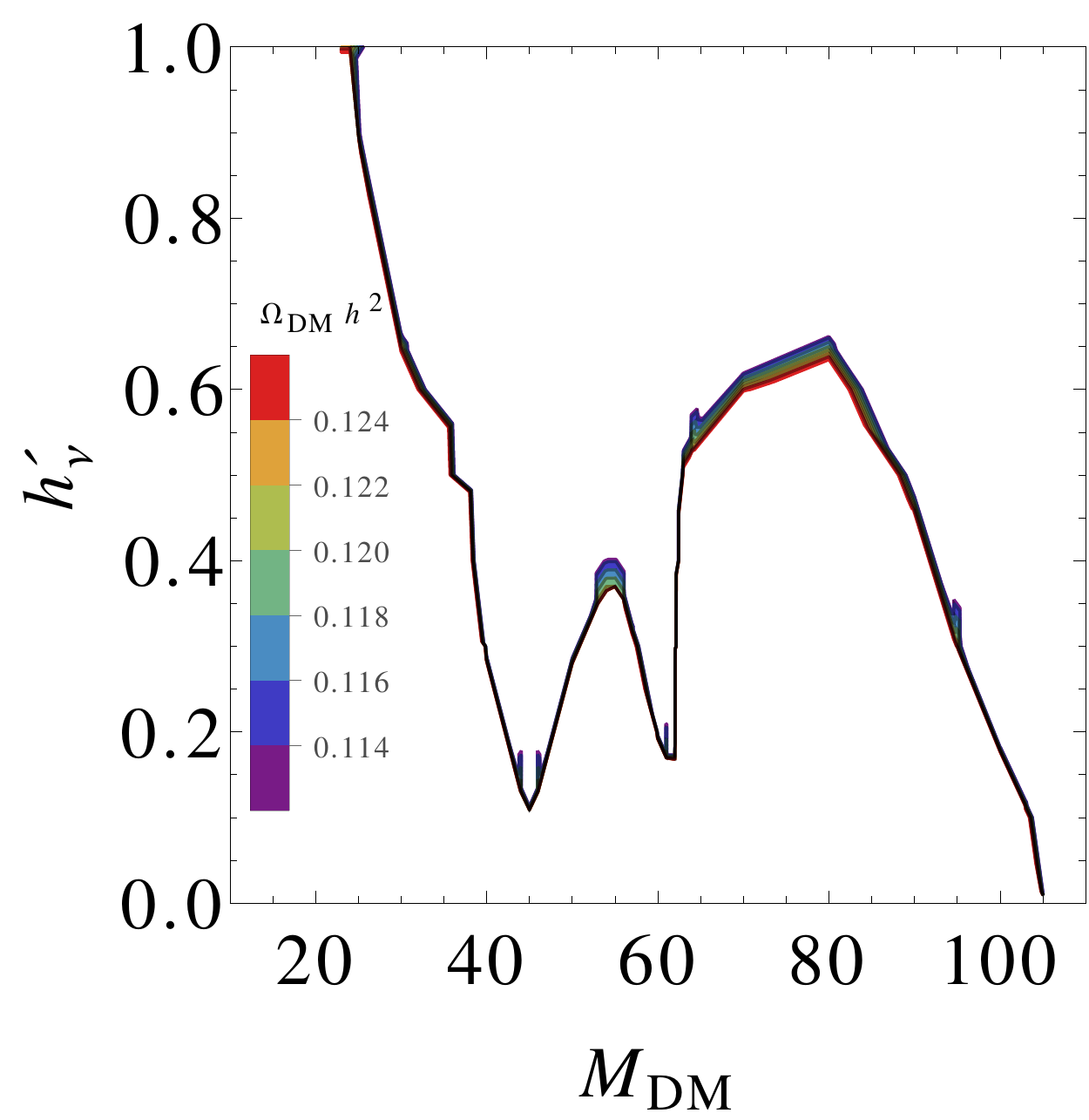}
\end{center}
\vskip -0.4in
\caption{(color online). \sl\small The relic density as a function of $M_{DM}$ (GeV) and  $h_\nu^\prime$.}
\label{RD}
\end{figure}

\section{Direct and Indirect Detection  }
\label{sec:DD}
%
 Direct detection offers the opportunity to detect DM as it passes through and scatters off normal matter.
The interaction can be classified as elastic or inelastic; and as spin-dependent or spin-independent.
In spin-dependent  scattering, the DM spin couples with the spin of the nucleon, while in spin-independent  scattering, the cross section does not depend on spin. 

\begin{figure}[htbp]
\vskip -0.3in
\begin{center}$
    \begin{array}{cc}
    \includegraphics[width=2.4in,height=2in]{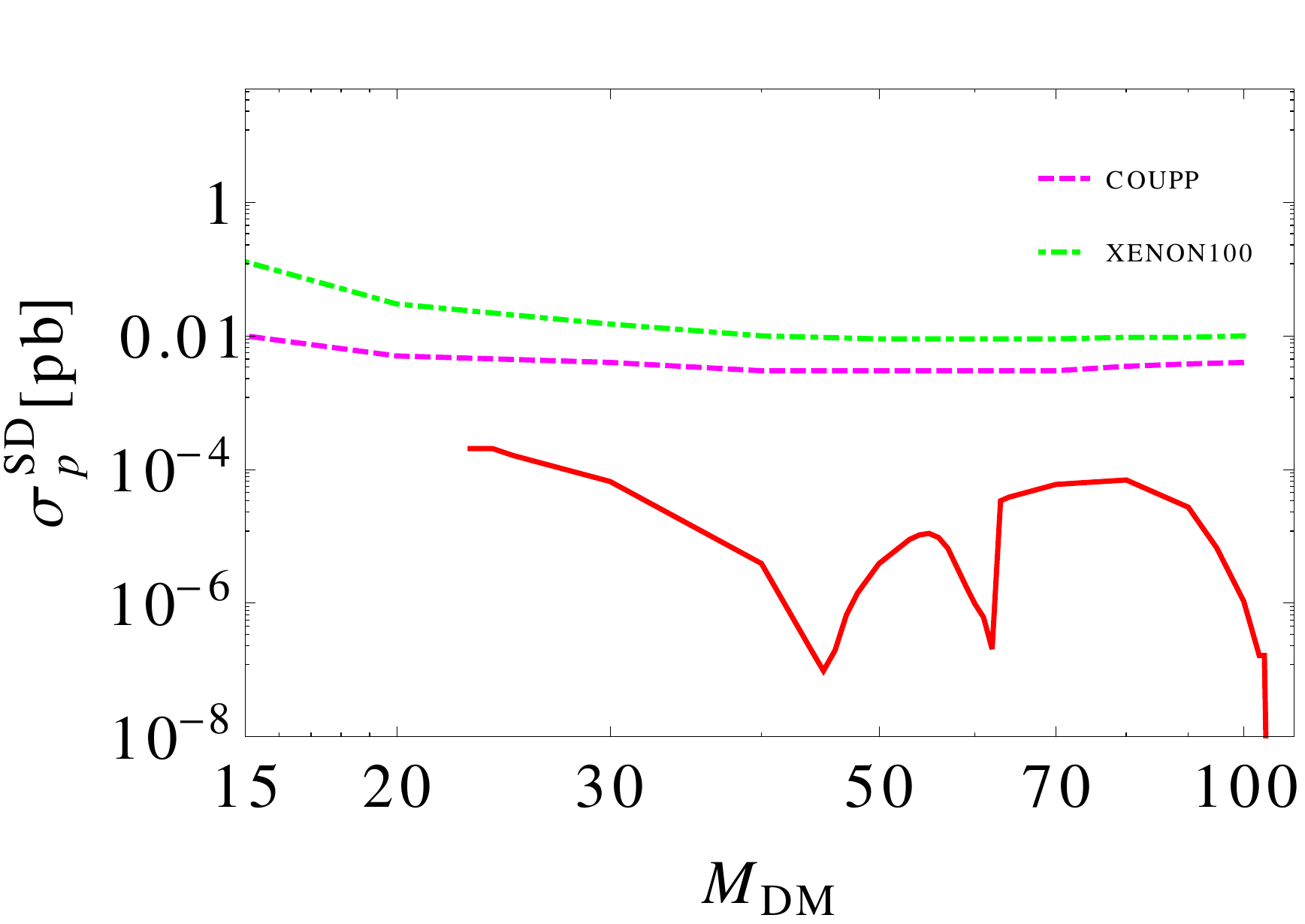}
&
    \includegraphics[width=2.4in,height=2in]{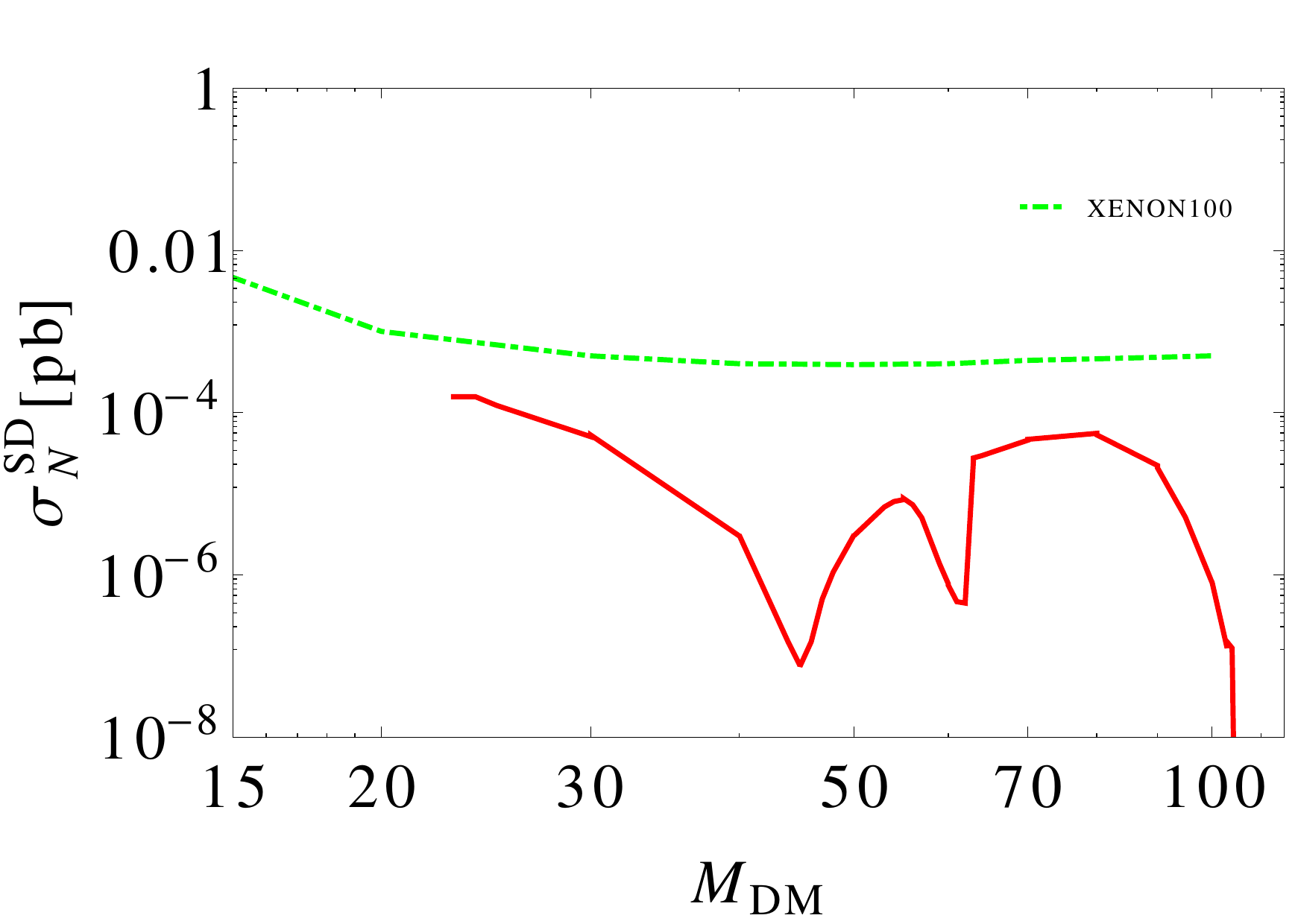}\\
        \includegraphics[width=2.5in,height=1.8in]{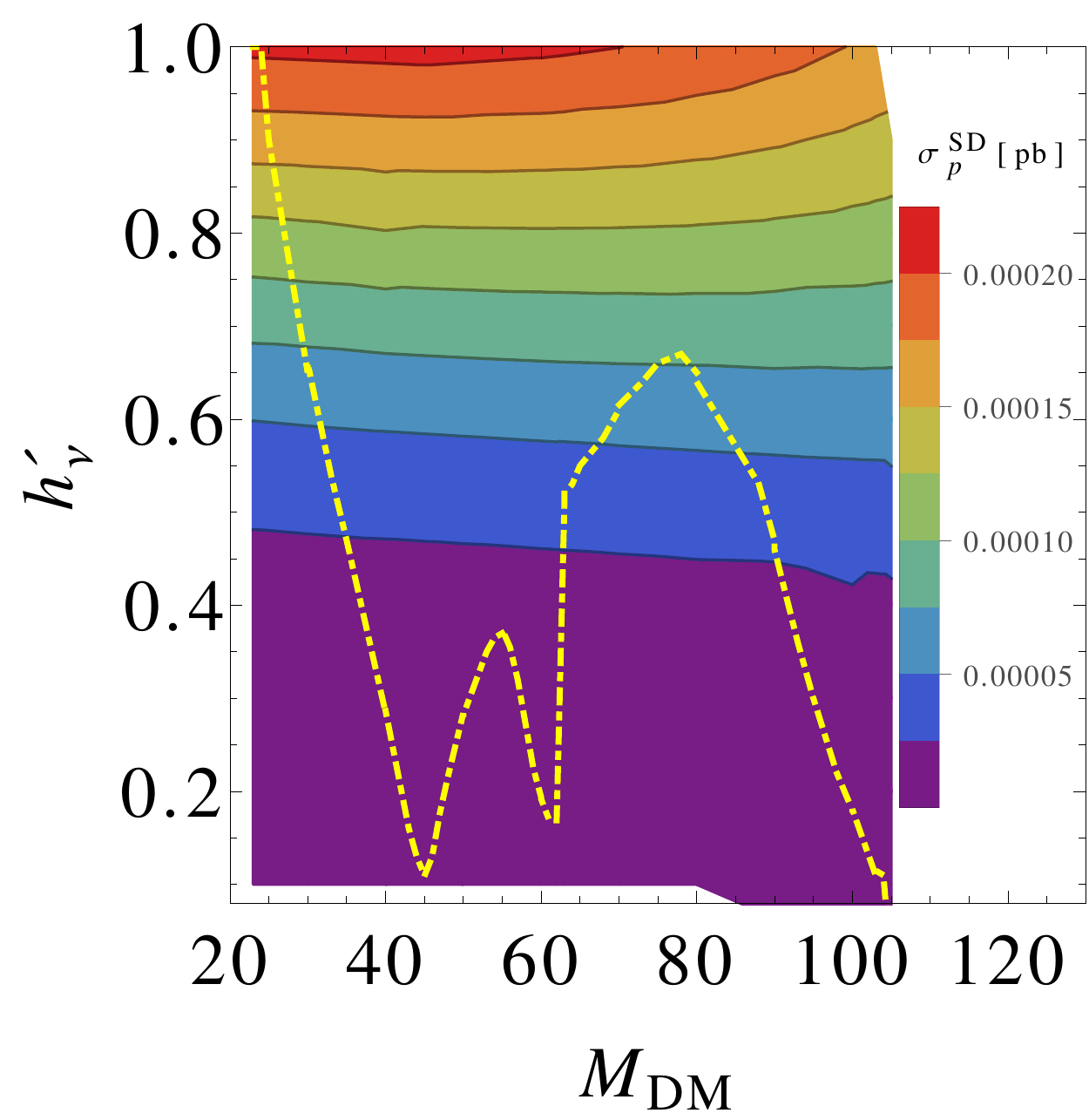}
&\hspace*{-0.2cm}
    \includegraphics[width=2.5in,height=1.8in]{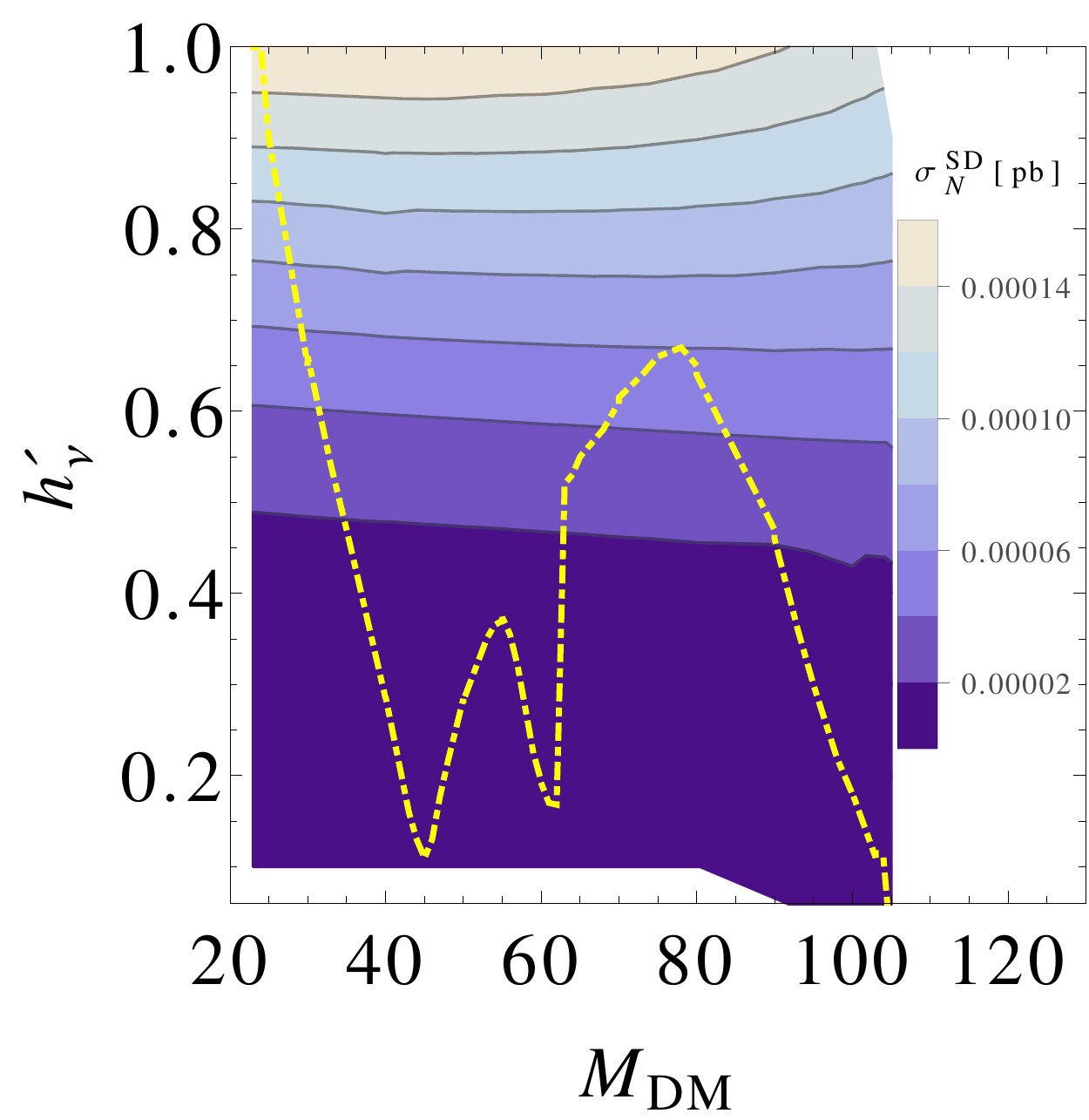}
        \end{array}$
\end{center}
\vskip -0.3in
    \caption{ (color online). \sl\small SD cross sections for proton (left) and neutron (right).}
\label{fig:SDlimit}
\end{figure}

Fig. \ref{fig:SDlimit}, upper panel, shows the spin-dependent (SD) cross section of DM, as a function of the DM mass $M_{DM}$, whereas  the bottom panel, shows the SD cross sections of the nucleon as contours in DM mass $M_{DM}$ and Yukawa coupling $h_\nu^\prime$ space.  Left panels are for the proton, the right ones for the neutron. The red lines show points of the parameter space which reproduce acceptable relic density. The areas above the pink dashed line and green dashed-dotted line are ruled out by the COUPP and XENON100  \cite{Behnke:2012ys, Aprile:2013doa} measurements, respectively. the experimental results do not restrict the parameter space of the model, but  only parameter points situated along the dash-dotted  yellow lines in the bottom panels give the correct dark matter relic density. 

\begin{figure}[htbp]
\vskip -0.3in
\begin{center}$
    \begin{array}{ccc}
	\hspace*{-1.1cm}\includegraphics[width=2.0in,height=2.0in]{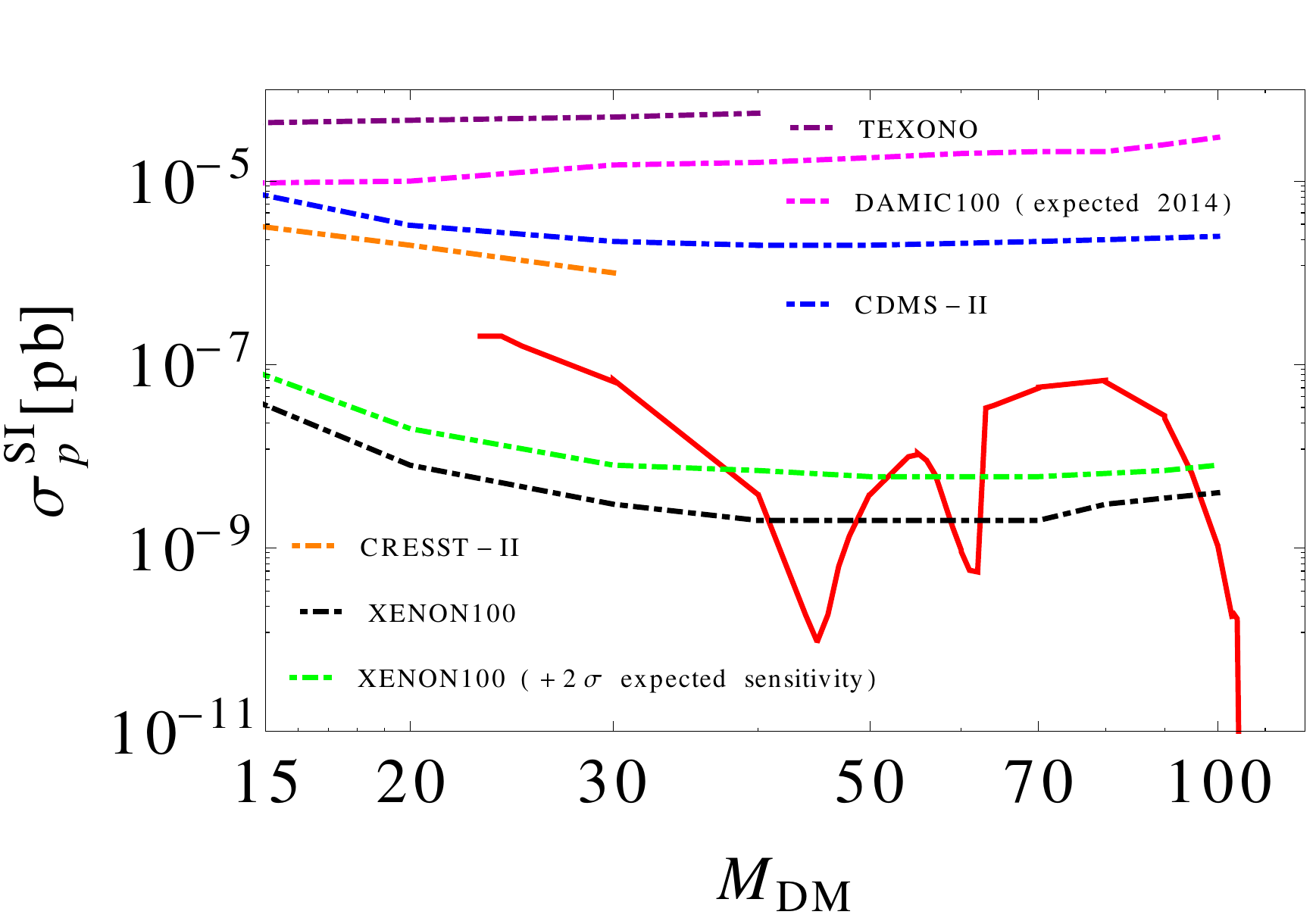}
	& \hspace*{-0.2cm}  \includegraphics[width=2.0in,height=1.8in]{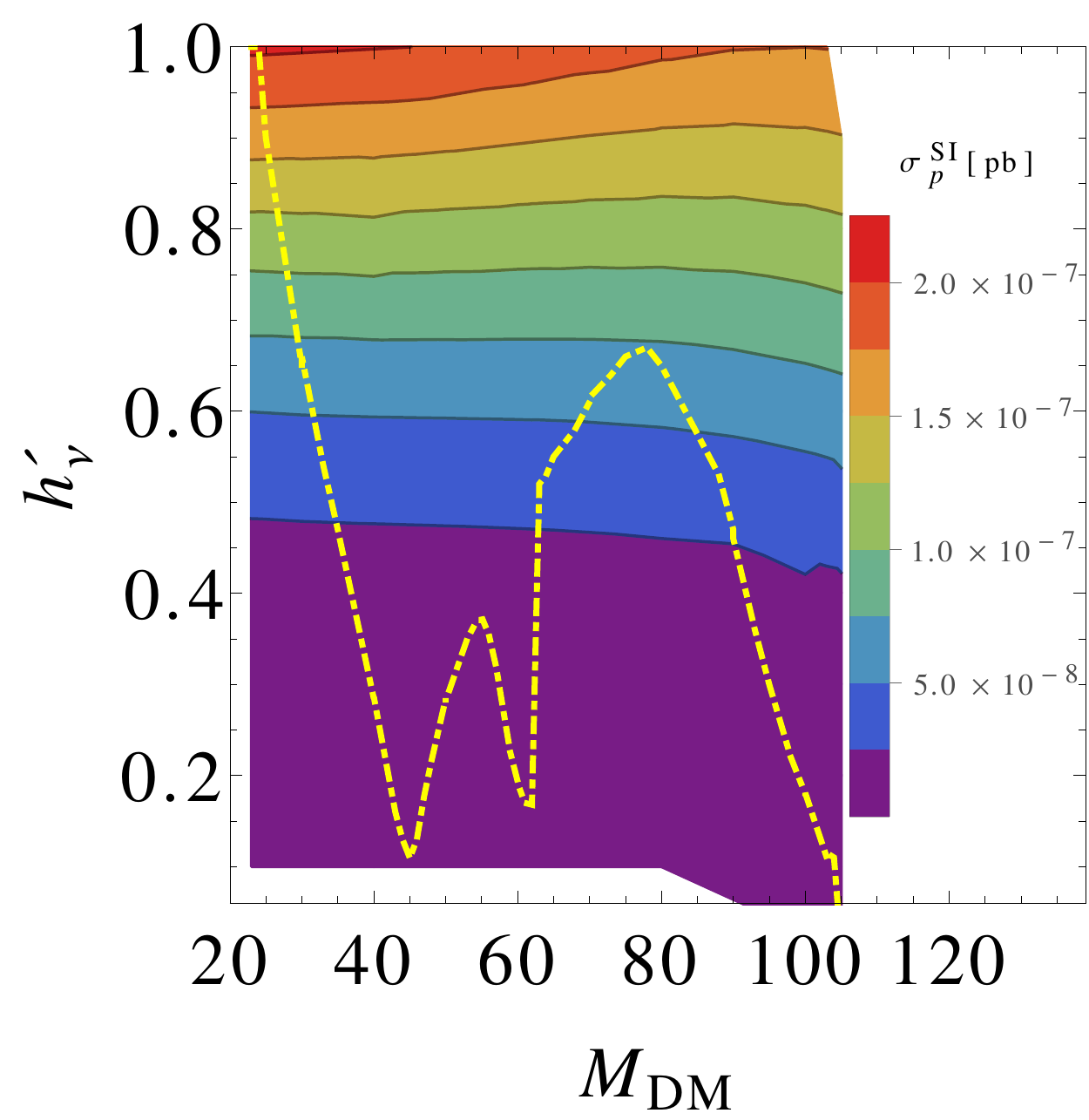}
	& \hspace*{-0.2cm}  \includegraphics[width=2.0in,height=1.8in]{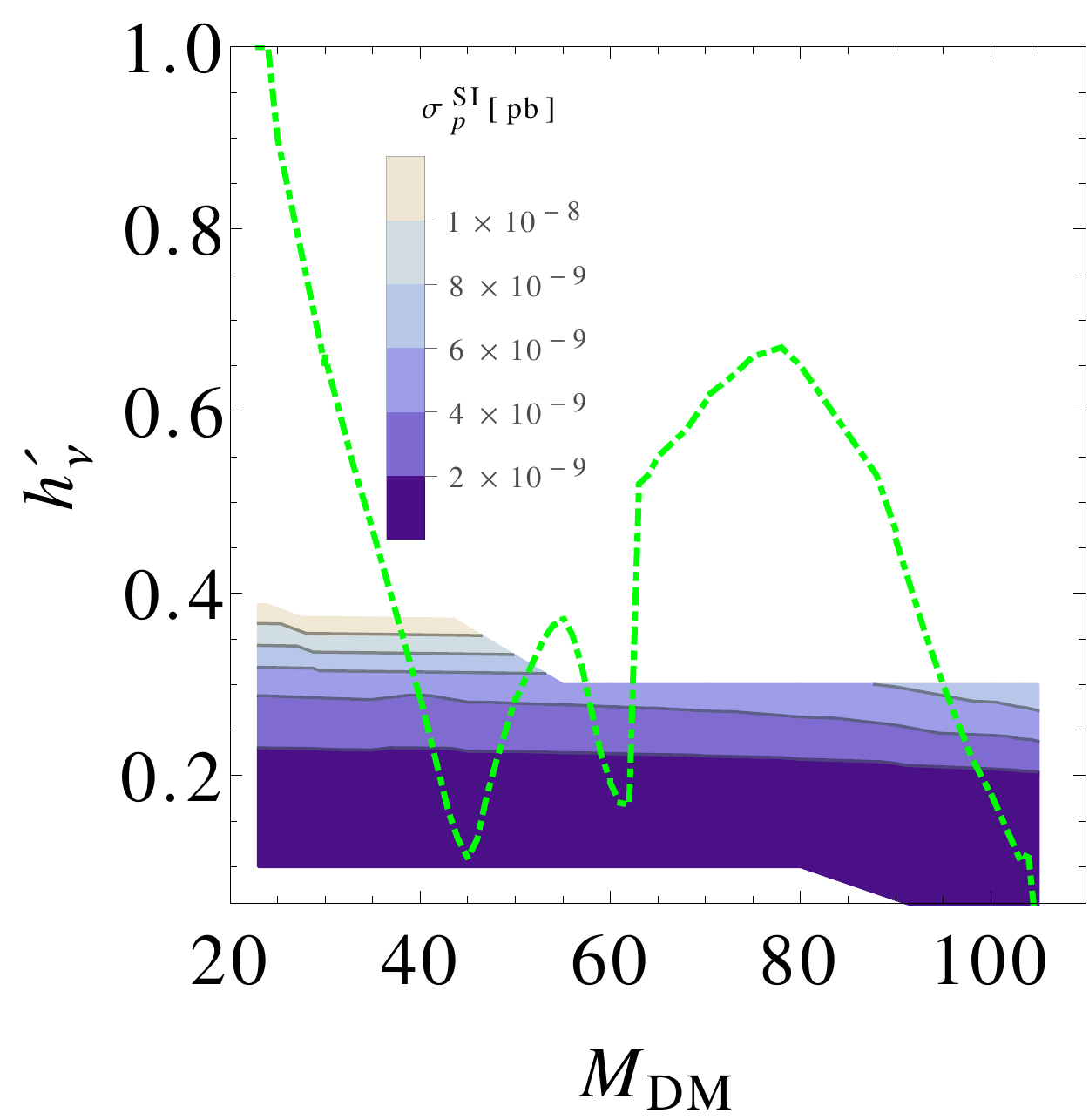}
        \end{array}$
\end{center}
\vskip -0.3in
    \caption{\sl\small SI cross sections, with all constraints (left) without XENON100 (middle) and with XENON100 (right).}
\label{fig:SI}
\end{figure}

In Fig. \ref{fig:SI} we show the spin-independent (SI) cross section of nucleon, as a function of the DM mass $M_{DM}$ (in GeV) (left  panel), the spin-independent cross section of the proton as a graph in $M_{DM}-h_\nu^\prime$ space,  constrained by all the experiments with the exception of XENON100 (middle panel), and including XENON100 (with $2\sigma$ expected sensitivity) measurements (right panel). The red/dashed line includes all points yielding consistent relic density. The regions above the dash-dotted black line, dash-dotted green line, dash-dotted orange line, dash-dotted blue line, dash-dotted purple line, dash-dotted pink line are ruled out by XENON100 \cite{Lavina:2013zxa}, XENON100 with $2\sigma$ expected sensitivity, CRESST-II \cite{Angloher:2014myn}, CDMS-II \cite{Agnese:2013cvt}, TEXONO \cite{Li:2013fla} and DAMIC100 (expected for 2014)  \cite{Chavarria:2014ika} results, respectively.  XENON100 results (with $2\sigma$ expected sensitivity) restrict the dark matter mass to be in the  37-52 GeV, or  57-63 GeV ranges, or heavier than 95 GeV, while white regions of parameter space are ruled out. 

\begin{figure}[htbp]
\vskip -0.2in
\begin{center}$
    \begin{array}{cc}
\hspace*{-0.5cm}
\includegraphics[width=2.2in,height=1.9in]{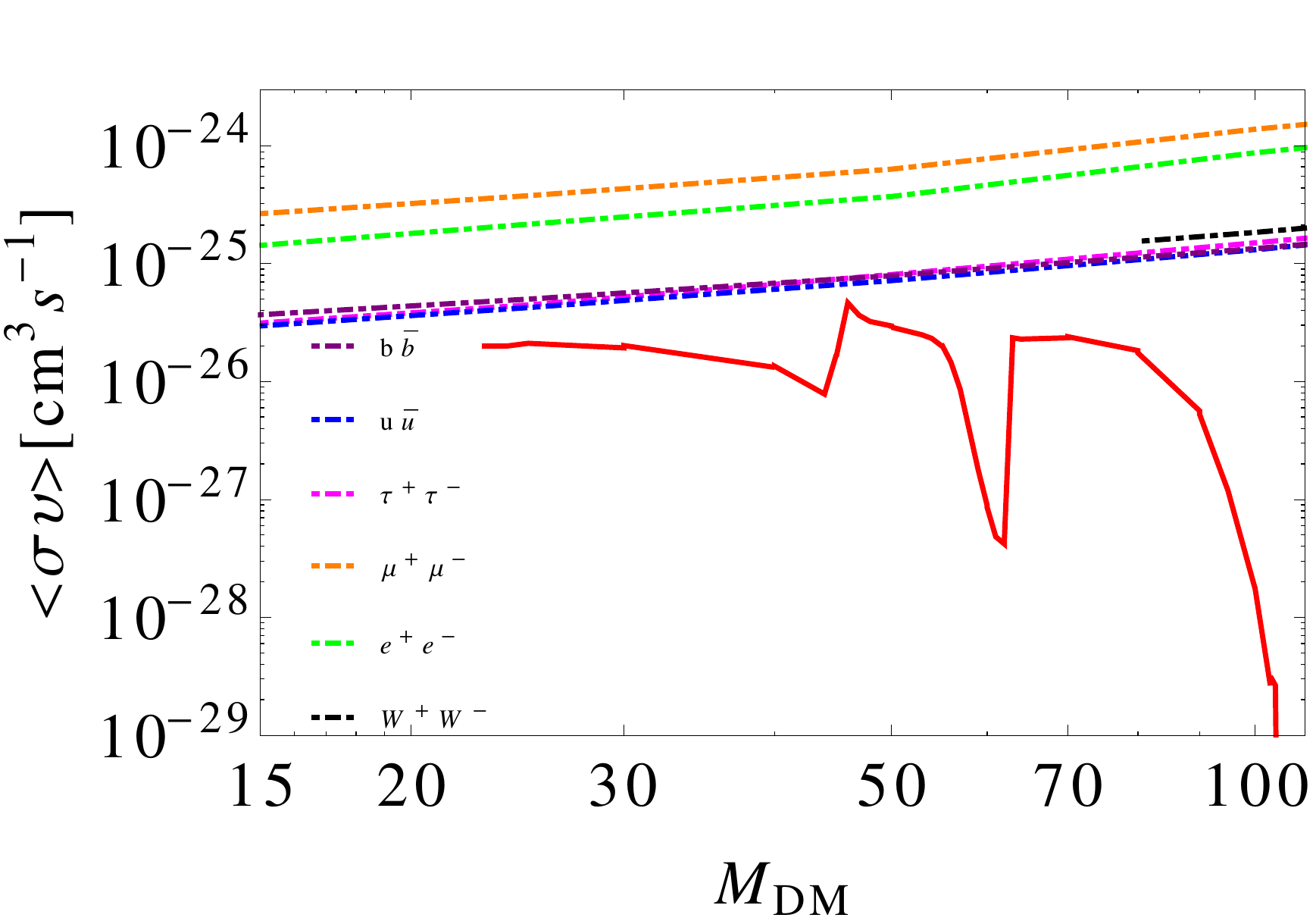}
    &\hspace*{-0.2cm}
   \includegraphics[width=2.4in,height=1.8in]{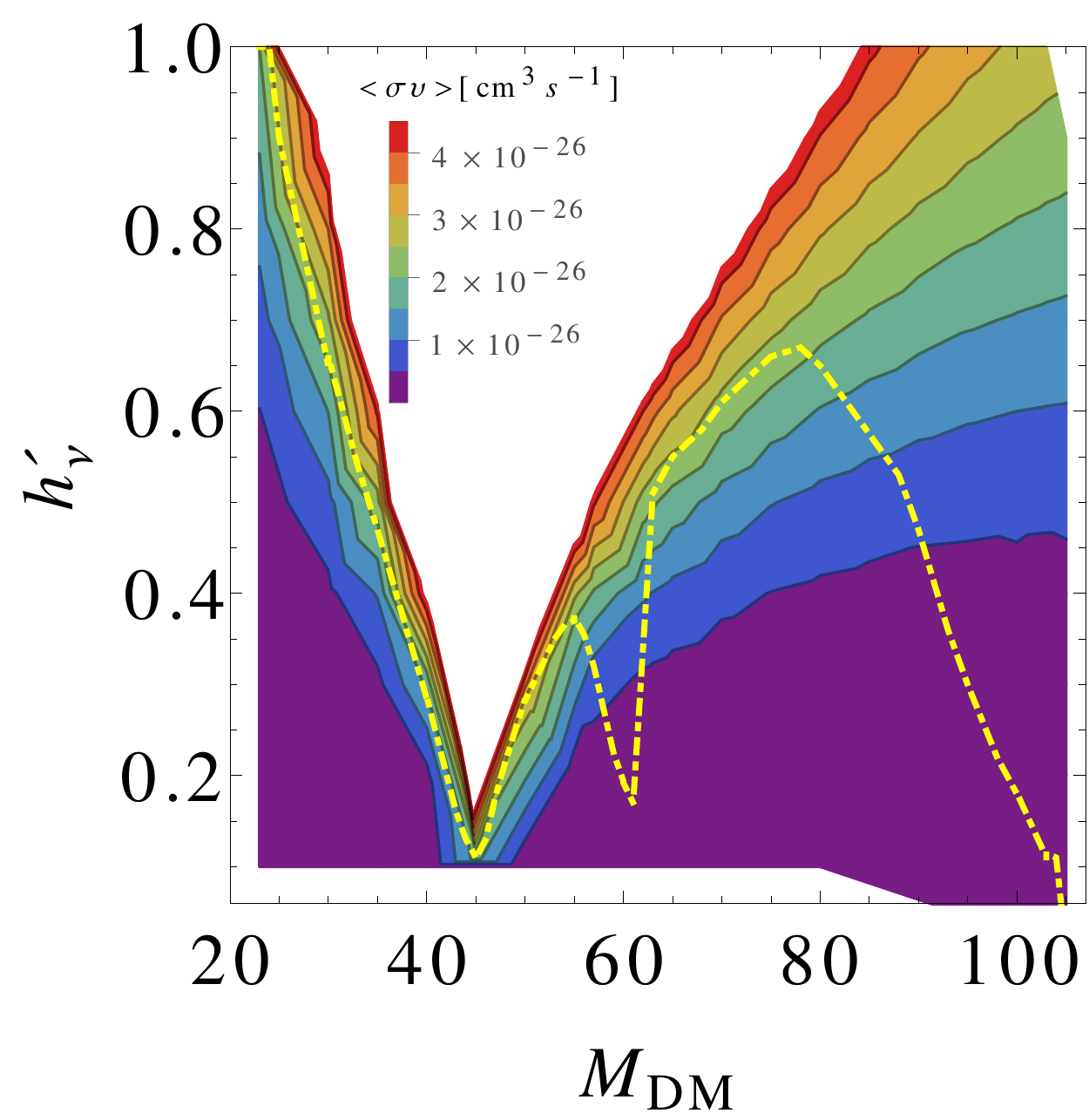}
  \end{array}$
\end{center}
\vskip -0.3in
\caption{\sl\small Annihilation cross  section of DM. }
\label{annihilation}
\end{figure}

 Indirect detection experiments for DM detection look for signatures of annihilations of DM particles in the flux of cosmic rays. 
In Fig. \ref{annihilation} we show the annihilation cross section of DM as a function of the DM mass $M_{DM}$,  compared with Fermi-LAT Collaboration results \cite{Ackermann:2013yva} (left panel); (right panel) contour plot for the annihilation cross section in the $M_{DM}$ - $h_\nu^\prime$ plane. The contours are consisted with the experimental values for the cross sections,  while the white regions are ruled out. Only  points along the dash-dotted yellow line give the correct dark matter relic density to 2$\sigma$. 

\begin{figure}[htbp]
\vskip -0.1in
\begin{center}$
    \begin{array}{cc}
    \includegraphics[width=2.1in,height=1.7in]{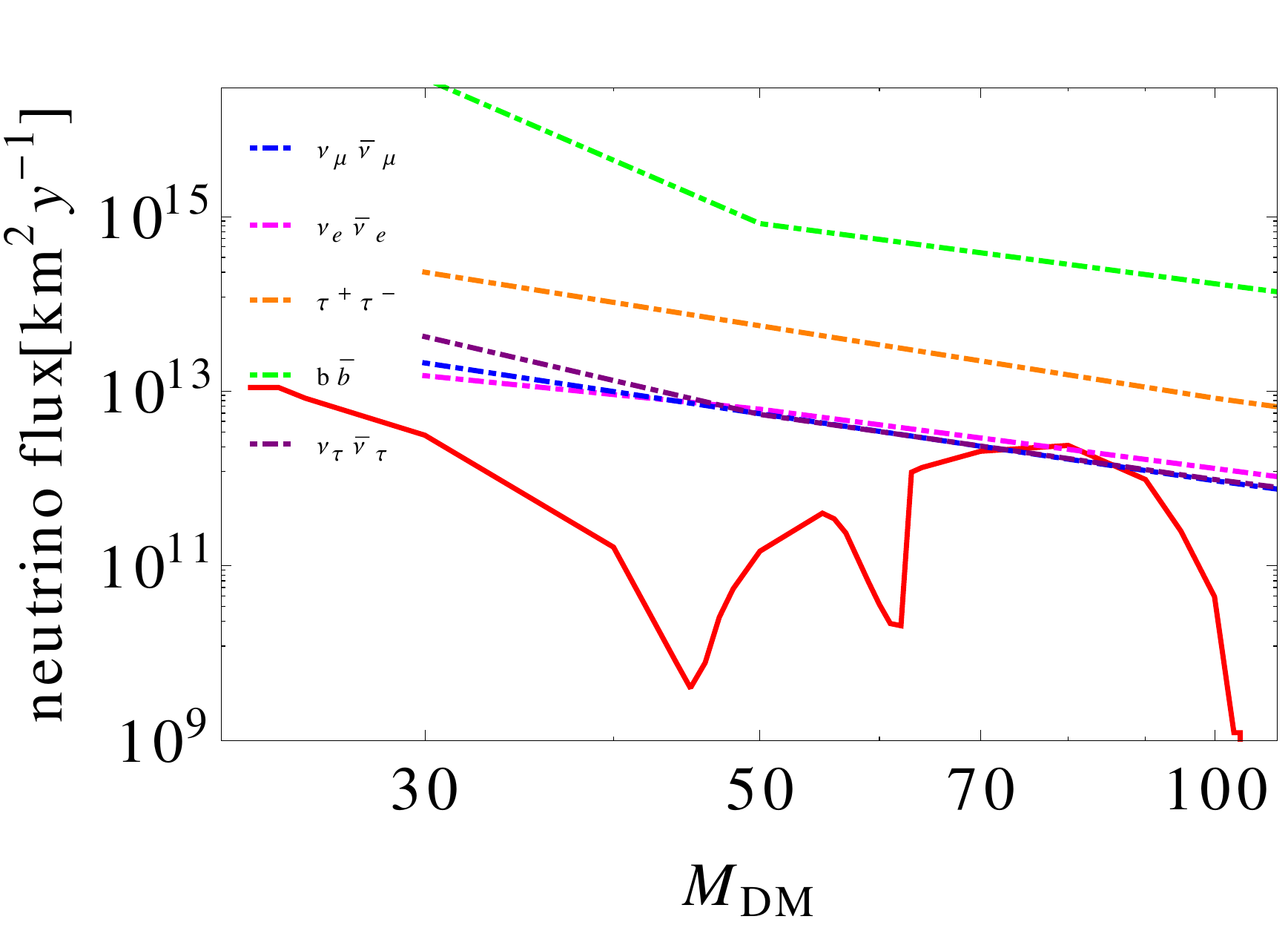}
&\hspace*{-0.1cm}
    \includegraphics[width=2.1in,height=1.7in]{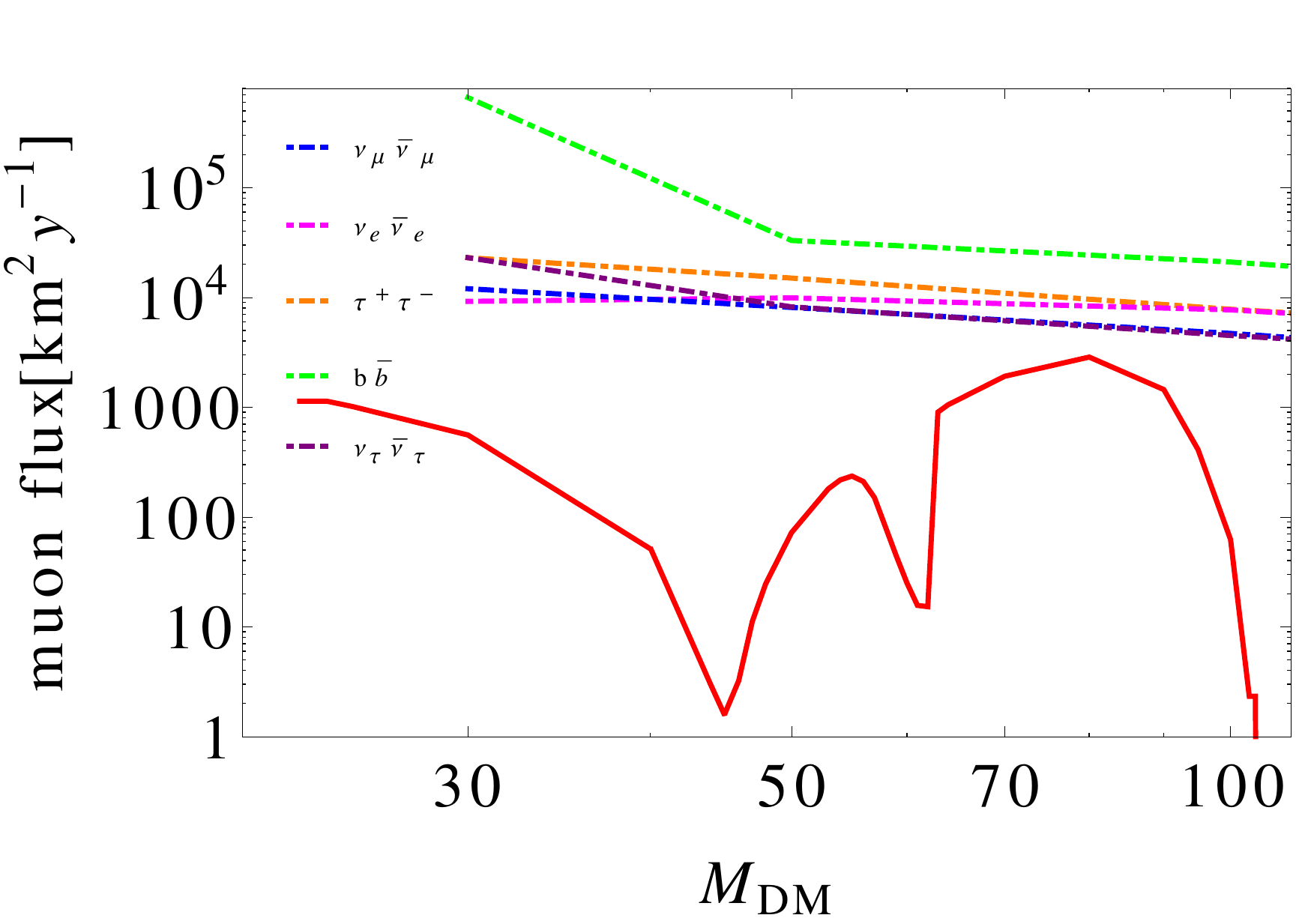}\\
           \includegraphics[width=2.3in,height=1.6in]{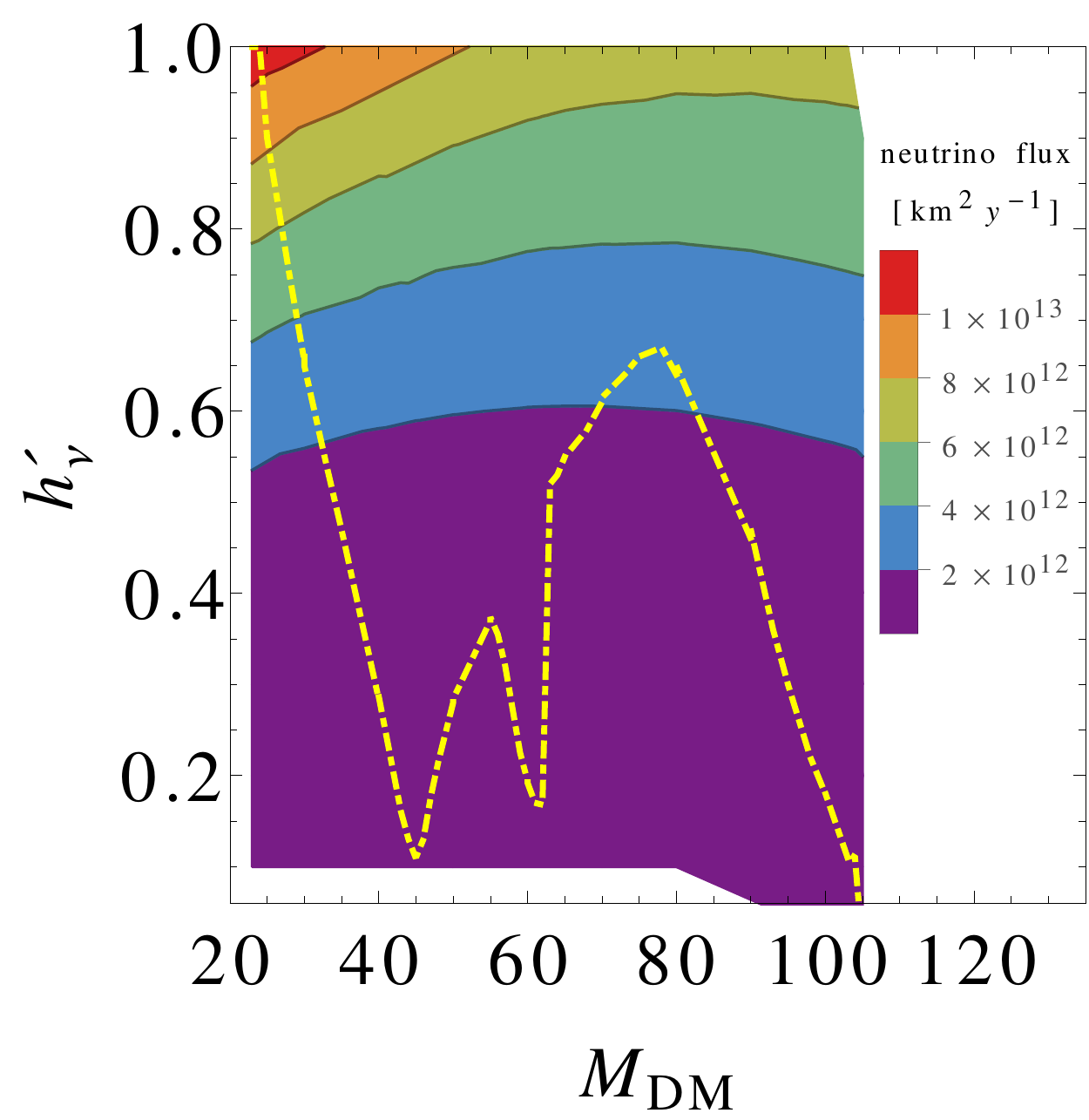}
&\hspace*{-0.2cm}
    \includegraphics[width=2.3in,height=1.6in]{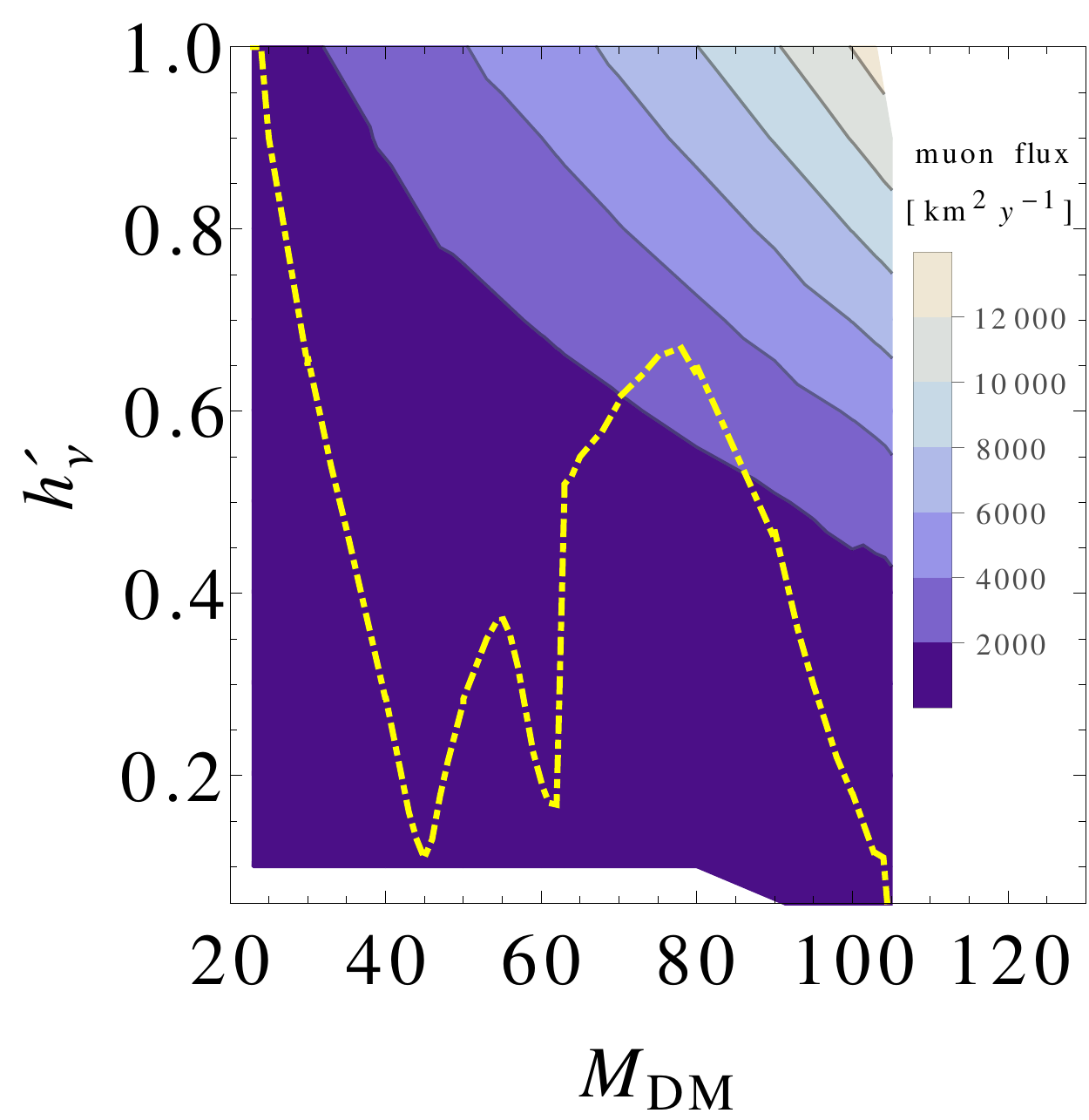}
        \end{array}$
\end{center}
\vskip -0.3in
     \caption{(color online). \sl\small Fluxes of neutrinos (left panel) and muons (right panel).}
\label{fig:NMfluxlimit}
\end{figure}

\vskip-0.1in
%
%
 Fig. \ref{fig:NMfluxlimit} shows the neutrino (left panel) and muon (right panel) fluxes as functions of $M_{DM}$ (in GeV). In the top graphs, our results are the red curve and the experimental results are from Baikal NT200 \cite{Avrorin:2014swy}. While the muon flux is consistent with all parameter points,  the neutrino flux excludes $M_{DM}$  in the 74-85 GeV region.  

\section{Conclusion}
\label{sec:conclusion}

To summarize, we  presented a  model with vector-like leptons that accounts for both neutrino masses and dark matter, and is consistent with the relic density and all direct and indirect searches. Assuming a single dark matter particle, the experimental data restricts the DM candidate to be light: in the 37-52 GeV, or 57-63 GeV range, or heavier than 95 GeV, all for points satisfying relic density constraints. In addition,  the neutrino flux excludes DM particles with mass in the 74-85 GeV range. 

\end{document}

%% file: econfmacros.tex
\def\beq{\begin{equation}}
\def\eeq#1{\label{#1}\end{equation}}
\def\eeqn{\end{equation}}

\def\beqa{\begin{eqnarray}}
\def\eeqa#1{\label{#1}\end{eqnarray}}
\def\eeqan{\end{eqnarray}}

\let\bar=\overbar

\def\Dslash{\not{\hbox{\kern-4pt $D$}}}
\def\dslash{\not{\hbox{\kern-2pt $\del$}}}

\def\msb{{\bar{\ssstyle M \kern -1pt S}}}




%% file: Neutrino_DM.bbl
\begin{thebibliography}{99}


\bibitem{Bahrami:2015mwa} 
  S.~Bahrami and M.~Frank,
  Phys.\ Rev.\ D {\bf 91}, 075003 (2015).
\bibitem{Komatsu:2010fb} 
  E.~Komatsu {\it et al.}  [WMAP Collaboration],
  Astrophys.\ J.\ Suppl.\  {\bf 192}, 18 (2011).

\bibitem{Ade:2013zuv} 
  P.~A.~R.~Ade {\it et al.}  [Planck Collaboration],
  Astron.\ Astrophys.\  {\bf 571}, A16 (2014).

\bibitem{Behnke:2012ys} 
  E.~Behnke {\it et al.}  [COUPP Collaboration],
  Phys.\ Rev.\ D {\bf 86}, no. 5, 052001 (2012)
  [Erratum-ibid.\ D {\bf 90}, no. 7, 079902 (2014)].

	
\bibitem{Aprile:2013doa} 
  E.~Aprile {\it et al.}  [XENON100 Collaboration],
  Phys.\ Rev.\ Lett.\  {\bf 111}, no. 2, 021301 (2013).
  
\bibitem{Lavina:2013zxa} 
  L.~S.~Lavina [ XENON100 Collaboration],
  arXiv:1305.0224 [hep-ex]; 
  E.~Aprile {\it et al.}  [XENON100 Collaboration],
  Phys.\ Rev.\ Lett.\  {\bf 109}, 181301 (2012);
  E.~Aprile {\it et al.}  [XENON100 Collaboration],
  Phys.\ Rev.\ Lett.\  {\bf 109}, 181301 (2012).

  
\bibitem{Angloher:2014myn} 
  G.~Angloher {\it et al.}  [CRESST-II Collaboration],
  arXiv:1407.3146 [astro-ph.CO].

  
  \bibitem{Agnese:2013cvt} 
  R.~Agnese {\it et al.}  [CDMS Collaboration],
  Phys.\ Rev.\ D {\bf 88}, 031104 (2013).
	
	
\bibitem{Li:2013fla} 
  H.~B.~Li {\it et al.}  [TEXONO Collaboration],
  Phys.\ Rev.\ Lett.\  {\bf 110}, no. 26, 261301 (2013).
	
\bibitem{Chavarria:2014ika} 
  A.~Chavarria, J.~Tiffenberg, A.~Aguilar-Arevalo, D.~Amidei, X.~Bertou, G.~Cancelo, J.~C.~D'Olivo and J.~Estrada {\it et al.},
  arXiv:1407.0347 [physics.ins-det].

  
\bibitem{Ackermann:2013yva} 
  M.~Ackermann {\it et al.}  [Fermi-LAT Collaboration],
  Phys.\ Rev.\ D {\bf 89}, no. 4, 042001 (2014).

	
\bibitem{Avrorin:2014swy} 
  A.~D.~Avrorin {\it et al.}  [Baikal Collaboration],
  Astroparticle Physics (2015), pp. 12-20.


\end{thebibliography}
